\documentclass[prd,
preprint,nofootinbib%
 ,secnumarabic%
,amssymb, amsmath,mathrsfs,nobibnotes,aps,11pt]{revtex4}
\usepackage{epsfig}
\usepackage{color}
\usepackage{ulem}
\usepackage{graphicx}
\usepackage{float}
\usepackage[caption=false]{subfig}
\usepackage{braket}
\usepackage{slashed}

\usepackage{tabularx}
\usepackage[colorlinks=true,citecolor=blue,linkcolor=blue]{hyperref}%
\begin{document}

\title{Very special relativity induced phase in neutrino oscillation  }
\author{Alekha C. Nayak\footnote{ alekha@prl.res.in}}
\affiliation{Physical Research Laboratory, Ahmedabad 380009, India }

\def\be{\begin{equation}}
\def\ee{\end{equation}}
\def\al{\alpha}
\def\bea{\begin{eqnarray}}
\def\eea{\end{eqnarray}}


\begin{abstract}
In Very Special Relativity (VSR), the neutrino mass term is coupled with the VSR preferred axis, and hence
Lorentz violating in nature. Beyond standard model physics predicts neutrino magnetic moment which
is linearly proportional to the mass eigenstates of the neutrinos. We report an additive kinematic phase in
the neutrino flavor oscillation due to the neutrino magnetic moment in the VSR framework. This phase is
proportional to the coupling between the VSR preferred axis with the external magnetic field as well as the spin of the
neutrino. Furthermore, we predict time variation in the neutrino oscillation with a period of one sidereal
day.
\end{abstract}
\maketitle


\section{Theory}
The particle interactions in the Standard Model (SM) are based on the assumption that it  preserves Lorentz and CPT symmetry. However, SM is a low energy limit of a unified theory at Planck scale. Local Lorentz and CPT symmetry violation are severely constrained by  the experiments\cite{Collins:2004bp,Jain:2005as,GrootNibbelink:2004za,Polchinski:2011za}.  To test these assumptions in experiments, an alternate theory of Lorentz violation have proposed by Cohen and Glashow\cite{Cohen:2006ky}. The null result of Michelson-Morley experiment does not necessarily require full Lorentz group,  it only demands particular subgroup of the Lorentz group such as T(2), E(2), HOM(2) and SIM(2)\cite{Cohen:2006ky}.  Adjoining space-time translation with these Lorentz subgroups in known as very special relativity. The generators for T(2) group are $T _1 = K _x + J_ y $ and $T _2 = K _y -J_ x$ , generator
of T(2) along with $J_ z $ forms E(2) subgroup, generator of T(2) along with $K _z$ forms HOM(2)
subgroup, generator of T(2) along with $K_ z$ and $J _z$ forms SIM(2) subgroup\cite{Cohen:2006ky}. Here $\bf{J}$ and $\bf{K}$ represent rotation and boost respectively.

The VSR invariant Lagrangian for neutrino is given by \cite{Cohen:2006ky,Cohen:2006ir}
\begin{eqnarray}
L=\bar{\nu}_L \big( i \slashed \partial -\frac{m_{\nu}^2}{2}\frac{\slashed n}{i n.\partial}\big)\nu_L \label{neutrino}
\end{eqnarray}
where the null vector: $n^{\mu}=(1,0,0,1)$ defines the preferred axis. $n^{\mu}$ is invariant under $T_1$, $T_2$ and $J_z$, but transforms under $K_z$ as : $n^{\mu} \rightarrow n^{\mu}  e^{\phi}$, where $e^\phi=(E-p_z)/m_{\nu}$  \cite{Cohen:2006ky,Cohen:2006ir}.
The presence of preferred axis $n^{\mu}$ in the second term of Eq.\eqref{neutrino} violates Lorentz symmetry. However, the numerator and denominator are homogeneous in $n^{\mu}$, hence it  becomes invariant under  HOM(2) and SIM(2) subgroup  \cite{Cohen:2006ky,Cohen:2006ir}. In principle, we can add such VSR
mass term for all the SM leptons by replacing the neutrino field with SM doublet and gauge it by
replacing $\partial_{\mu}$ with covariant derivative $D_{\mu}$. Diagonalization of the Yukawa mass term and VSR
mass term by making bi-unitary transformations, we get oscillation in the neutrino sector as well as in the charge lepton sector\cite{Dunn:2006xk,Nayak:2016zed}.

In the Pontecorvo framework of neutrino oscillation, the flavor eigenstates (which are produced in the weak  interaction) can be written as a linear superposition of mass eigenstates,
\bea
| \nu_{\ell}\rangle =\sum_{\ell} U_{\ell j}^{\ast} \,| \nu_{j}\rangle
\eea
where $| \nu_{j}\rangle$ is the mass eigenstate of neutrino. The probability of transition from $\nu_{\ell} \rightarrow \nu_{\ell^{\prime}}$ is given by
\bea
P(\nu_{\ell} \rightarrow \nu_{\ell^{\prime}
}) = \Big |\sum_{j\neq k} U_{\ell^{\prime}j }\,e^{\frac{-\Delta m_{kj}^2\, L}{2 E}}\, U_{\ell j}^{\ast} \Big|^2
\eea 
where $U_{\ell^{\prime}j }$ is the probability  amplitude to find the state $| \nu_{j}\rangle$ in flavor state $| \nu_{\ell^{\prime}}\rangle$,  $E$ is the energy of the neutrino beam and L is the distance between source and detector. The kinematically phase in the neutrino oscillation is given by 
\bea
\phi_{k j}= \frac{\Delta m_{kj}^2\, L}{2 E}\,=\,1.27 \,\Big(\frac{\Delta m_{kj}^2}{eV^2} \Big)\, \Big( \frac{L}{km}\Big)\,\Big( \frac{GeV}{E}\Big)
\label{standard}
\eea

In this letter, we report an additive VSR induced phase to the QFT induced phase\cite{Ahluwalia:2017rsu} due to the magnetic moment of neutrino. The modified VSR invariant Dirac equation can be written as 
\bea
\Big(\slashed {p}-m_{\nu}^2\,\frac{\slashed{n}}{2 n.p}\Big)\, \nu_L=0
\label{eq:dirac}
\eea
Applying the modified Dirac operator from the left-hand side of Eq.\eqref{eq:dirac}, we get $p^2=m_{\nu}^2$. Since the the dispersion relation remain unchanged, the free particle propagation does not change in VSR. 

The modified Dirac spinor for positive energy solution can be written as \cite{Dunn:2006xk}
\bea
u^{\prime}(p) \approx \Big(1-\frac{m_{\nu}^2}{4}\frac{\slashed {n}}{n.p}  \Big)u(p)
\eea
where $u(p)$ is the positive energy Dirac spinor.  
The  matrix element for neutrino interacting with the electromagnetic field can be expressed as : $i\bar{u}^{\prime}(p^{\prime},s^{\prime}) \, \Gamma^{\mu}_{s,\,s^{\prime}}(p,\,p^{\prime})\, u^{\prime}(p,s)A_{\mu}$. In the $n.A=0$ gauge, the effective interaction of the neutrino with electromagnetic field due to its magnetic moment  can be written as \cite{Dunn:2006xk}
\bea
i\bar{u}^{\prime}(p^{\prime},s^{\prime}) \, \Gamma^{\mu}_{s,\,s^{\prime}}(p,\,p^{\prime})\, u^{\prime}(p,s)A_{\mu} \approx i \mathcal{\mu}_{\nu}\, \frac{m_{\nu}^2}{2} \, \bar{u}(p^{\prime},s^{\prime}) \, \frac{n_{\nu}\,n^{\beta}\,\tilde{F}^{\mu \nu} \sigma_{\beta \nu} }{(n.p)\, (n.p^{\prime})}\, u(p,s)
\label{eq:int}
\eea

In the non-relativistic limit, the interaction term can be expanded as 
\bea
i \mathcal{\mu}_{\nu}\, \frac{m_{\nu}^2}{2} \, \bar{u}(p^{\prime},s^{\prime}) \, \frac{n_{\nu}\,n^{\beta}\,\tilde{F}^{\mu \nu} \sigma_{\beta \nu} }{(n.p)\, (n.p^{\prime})}\, u(p,s) \approx i \mathcal{\mu}_{\nu} \, \chi^{\dagger} [(\bf{n}. \boldsymbol{\sigma})(\bf{n}.\bf{B})- \boldsymbol{\sigma}.\bf{B}+(\bf{n}\times\bf{E}). \boldsymbol{\sigma}]\chi
\label{eq:int2}
\eea

The corresponding interaction Hamiltonian can be written as \cite{Dunn:2006xk,Fan:2006nd}
\bea
H_{VSR}= \mathcal{\mu}_{\nu}\, [ \boldsymbol{\sigma}.\bf{B} -(\bf{n}. \boldsymbol{\sigma})(\bf{n}.\bf{B})-(\bf{n}\times\bf{E}). \boldsymbol{\sigma}]
\label{eq:ham}
\eea
Neutrino magnetic moment has been studied in the beyond standard model physics with nonzero neutrino mass\cite{Marciano:1977wx,Lee:1977tib,Pal:1981rm,Schechter:1981hw,Shrock:1982sc,Masood:1992dg,Masood:1995aq,Dvornikov:2004sj,Bell:2005kz}. The neutrino magnetic moment due to quantum correction is proportional to its mass eigenstate  \cite{Marciano:1977wx,Lee:1977tib}
 \bea
\mathcal{\mu}_{\nu\,i}= \frac{3 e G_f m_{\nu \,i}}{8 \sqrt{2}\pi^2}
 \eea 
We can drop the third term in Eq.\eqref{eq:ham}, because the Earth and compact objects like neutron stars do not have its own electric field. The additive phase due to $\mathcal{\mu}_{\nu}\, \boldsymbol{\sigma}.\bf{B}$  in the flavor oscillation has been reported in Ref.\cite{Ahluwalia:2017rsu}. The new additive  phase to the kinematic phase ($\phi_{kj}$) in the neutrino flavor oscillation due to  the neutrino magnetic moment in the VSR framework can be written as 
\bea
\phi^{new}&=&\phi_{kj}\,\phi^{VSR} \nonumber \\
&=& - \phi_{kj}\, \mathcal{\mu}_{\nu}\,  (\bf{n}. \boldsymbol{\sigma})(\bf{n}.\bf{B})
\label{eq:ham}
\eea
where $\phi^{VSR}=-\mathcal{\mu}_{\nu}\,  (\bf{n}. \boldsymbol{\sigma})(\bf{n}.\bf{B})$. The new phase $\phi^{VSR}$  depends upon the angle between the  VSR preferred direction with the spin of the neutrino as well as the external magnetic field.   Furthermore,   Earth rotation changes the angle between the VSR preferred axis and neutrino beam, hence we may observe a change in the probability amplitude of conversion from one flavor to another in the neutrino oscillation data. This phase is also linearly proportional to the neutrino masse. In the next section, we will discuss about time variation in the   neutrino oscillation in one sidereal day due to this VSR phase. 

\section{Daily variation in the neutrino Oscillation data due to VSR phase}
Let us assume the observer is at latitude $\lambda$ in local laboratory coordinate system $XYZ$ and the astronomical equatorial system is denoted by $xyz$. The Earth rotation axis is parallel to the $z$ axis of the equatorial system. Let the unit vector along the $x,\,y,\,z$ axis be  $\hat{x},\, \hat{y},\, \hat{z}$ respectively and unit vector along $X,\, Y, \, Z$ axis be $\hat{X},\,\hat{Y}\, \hat{Z}$ respectively. We take $\hat{x}$ towards vernal equinox,  $\hat{Z}$ is vertically upward in the local frame, $\hat{X}$  and $\hat{Y}$ vectors are tangential to the surface pointing towards north and west respectively. Let $\beta$ is the right ascension of $\hat{j}$ at initial time t=0.  These two coordinates are related by 
\bea
&&\hat{x}=\cos(\theta+\beta)\,\hat{Y}-\sin(\theta+\beta)(\cos \lambda\, \hat{Z}-\sin \lambda \,\hat{X}) \nonumber \\
&&\hat{y}=\cos(\theta+\beta)\,\hat{Y}+\cos(\theta+\beta)(\cos \lambda\, \hat{Z}-\sin \lambda \,\hat{X}) \nonumber \\
&&\hat{z}= \cos \lambda \,\hat{X}+\sin \lambda \,\hat{Z}
\eea

The preferred direction $\bf{n}$ makes an angle $(\theta_e,\,\phi_e)$ in the equatorial coordinate system, where $\theta_e$  and $\phi_e$ are polar angle and right ascension.  The vector $\bf{n}$ in the local laboratory system is given by 
\bea
\hat{n}&=& [ \sin \lambda\, \sin \theta_e \,\sin (\theta+\beta-\phi_e)+\cos\lambda \,\cos\theta_e]\,\hat{X} + [\sin\theta_e\,\cos(\theta+\beta-\phi_e)] \,\hat{Y} \nonumber \\
&+&[\cos\theta_e\,\sin\lambda-\sin\theta_e\cos\lambda\,\sin(\theta+\beta-\phi_e)]\,\hat{Z}
\eea
where $\theta= \frac{2 \pi}{T_0}t$, $T_0$ is the one sidereal day. Let us take the magnetic field in the $\hat{Z}$ direction, i.e. ${\bf{B}}=B_0 \,\hat{Z}$ and the spin orientation of neutrino in the laboratory frame is given by ${ \boldsymbol{\sigma}}= \sigma_1\, \hat{X}+\sigma_2\, \hat{Y} + \sigma_3\, \hat{Z}$. 
The quantum mechanical operator acts on the spin state $|p, \sigma \rangle$ as follows: ${\bf{n}}. \boldsymbol{ \sigma}|p, \sigma \rangle = s_j|p, \sigma \rangle$, where $s_j$ is the eigenvalue which is equals to $\pm 1$.  Then, the new phase due to VSR is given by 
 \bea
&&\phi^{VSR}= - \mathcal{\mu}_{\nu}\,  ({\bf{n}}.{ \boldsymbol{\sigma}})({\bf{n}.\bf{B}}) \nonumber \\
&=& \pm  \mathcal{\mu}_{\nu}\, B_0 [\cos\theta_e\,\sin\lambda-\sin\theta_e\cos\lambda\,\sin(\theta+\beta-\phi_e)]  \nonumber \\
&=& \pm 3.7\times 10^{-27} \frac{B_0}{\text{Gauss}}[\cos\theta_e\,\sin\lambda-\sin\theta_e\cos\lambda\,\sin(\theta+\beta-\phi_e)] 
\eea
\begin{figure}[t]
  \begin{minipage}[]{0.50\linewidth}
  \centering
    \includegraphics[width=\linewidth]{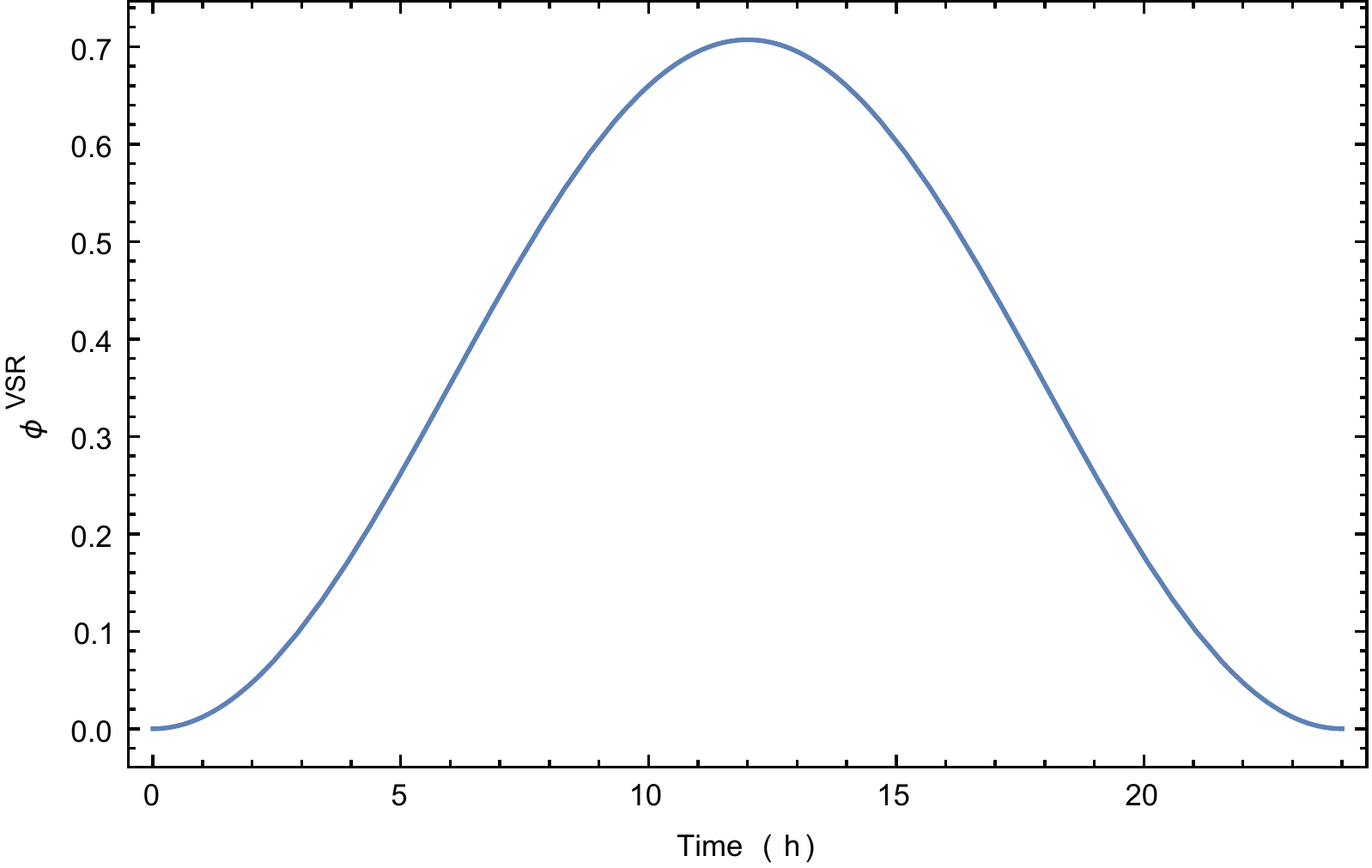}
  \end{minipage}
  \caption{The time  variation  of  the VSR phase ($\phi^{VSR}$) in an arbitrary unit for one sidereal day.  Here we choose   $ \theta_e=\pi/4, \, \beta-\phi_e=\pi/2 $ and latitude $\lambda=\pi/3$. } 
  \label{time_var}
\end{figure}
Due to  the rotation of the Earth, the angle between the  preferred axis and magnetic field changes with time (t), so the neutrino oscillation phase show time variation. Hence, the conversion probability of neutrino from one flavor to another also show time variation in the laboratory coordinate system. The time variation of $\phi^{VSR}$ is shown in Fig.\eqref{time_var} and time period  of this phase is  one sidereal day. 

\section{SUMMARY AND CONCLUSIONS}
In the VSR, an additional  kinematic phase is induced in the neutrino flavor oscillation  due to the  neutrino magnetic moment. This phase show time dependence in the neutrino flavor oscillation with a time period of one sidereal day. Since Earth magnetic field is $\sim$ 0.25 to 0.65 Gauss, the VSR induced oscillation is roughly twenty seven order of magnitude smaller than standard flavor oscillation (i.e. Eq.\eqref{standard}). For some magnetic white dwarf and magnetar the magnetic field strength is order of $10^{9}$ Gauss\cite{Hu:2018lwv} and $10^{16}$ Gauss\cite{Tiengo:2013gsa} respectively, hence this VSR induced oscillation is  eighteen and eleven orders of magnitude smaller than standard flavor oscillation in these compact objects.  Large neutrino magnetic moment in the order of $10^{-11} \mu_B$ ($\sim 10^{-20}\, eV/Gauss $) has been proposed\cite{PhysRevLett.65.2626,PhysRevD.46.2268,Lindner:2017uvt}. Such large neutrino magnetic moment enhances the $\phi^{VSR}$   effect by eight orders of magnitude in all the above scenario and that could bring the effect to possible observability. If  neutrino mass has  VSR origin, then this novel feature of time variation in the neutrino oscillation is expected.  


\textit{Acknowledgements:} I would like to thank Prof. Dharam Vir Ahluwalia for discussions and acknowledges the hospitality provided by IUCAA, Pune, India, during my visit where the work has been initiated. I also thank my
friends and colleagues Tripurari Srivastava and Arpan Das for reading
this draft and giving useful suggestions.


\bibliography{vdmref}

\begin{thebibliography}{24}
\expandafter\ifx\csname natexlab\endcsname\relax\def\natexlab#1{#1}\fi
\expandafter\ifx\csname bibnamefont\endcsname\relax
  \def\bibnamefont#1{#1}\fi
\expandafter\ifx\csname bibfnamefont\endcsname\relax
  \def\bibfnamefont#1{#1}\fi
\expandafter\ifx\csname citenamefont\endcsname\relax
  \def\citenamefont#1{#1}\fi
\expandafter\ifx\csname url\endcsname\relax
  \def\url#1{\texttt{#1}}\fi
\expandafter\ifx\csname urlprefix\endcsname\relax\def\urlprefix{URL }\fi
\providecommand{\bibinfo}[2]{#2}
\providecommand{\eprint}[2][]{\url{#2}}

\bibitem[{\citenamefont{Collins et~al.}(2004)\citenamefont{Collins, Perez,
  Sudarsky, Urrutia, and Vucetich}}]{Collins:2004bp}
\bibinfo{author}{\bibfnamefont{J.}~\bibnamefont{Collins}},
  \bibinfo{author}{\bibfnamefont{A.}~\bibnamefont{Perez}},
  \bibinfo{author}{\bibfnamefont{D.}~\bibnamefont{Sudarsky}},
  \bibinfo{author}{\bibfnamefont{L.}~\bibnamefont{Urrutia}}, \bibnamefont{and}
  \bibinfo{author}{\bibfnamefont{H.}~\bibnamefont{Vucetich}},
  \bibinfo{journal}{Phys. Rev. Lett.} \textbf{\bibinfo{volume}{93}},
  \bibinfo{pages}{191301} (\bibinfo{year}{2004}), \eprint{gr-qc/0403053}.

\bibitem[{\citenamefont{Jain and Ralston}(2005)}]{Jain:2005as}
\bibinfo{author}{\bibfnamefont{P.}~\bibnamefont{Jain}} \bibnamefont{and}
  \bibinfo{author}{\bibfnamefont{J.~P.} \bibnamefont{Ralston}},
  \bibinfo{journal}{Phys. Lett.} \textbf{\bibinfo{volume}{B621}},
  \bibinfo{pages}{213} (\bibinfo{year}{2005}), \eprint{hep-ph/0502106}.

\bibitem[{\citenamefont{Groot~Nibbelink and
  Pospelov}(2005)}]{GrootNibbelink:2004za}
\bibinfo{author}{\bibfnamefont{S.}~\bibnamefont{Groot~Nibbelink}}
  \bibnamefont{and} \bibinfo{author}{\bibfnamefont{M.}~\bibnamefont{Pospelov}},
  \bibinfo{journal}{Phys. Rev. Lett.} \textbf{\bibinfo{volume}{94}},
  \bibinfo{pages}{081601} (\bibinfo{year}{2005}), \eprint{hep-ph/0404271}.

\bibitem[{\citenamefont{Polchinski}(2012)}]{Polchinski:2011za}
\bibinfo{author}{\bibfnamefont{J.}~\bibnamefont{Polchinski}},
  \bibinfo{journal}{Class. Quant. Grav.} \textbf{\bibinfo{volume}{29}},
  \bibinfo{pages}{088001} (\bibinfo{year}{2012}), \eprint{1106.6346}.

\bibitem[{\citenamefont{Cohen and Glashow}(2006{\natexlab{a}})}]{Cohen:2006ky}
\bibinfo{author}{\bibfnamefont{A.~G.} \bibnamefont{Cohen}} \bibnamefont{and}
  \bibinfo{author}{\bibfnamefont{S.~L.} \bibnamefont{Glashow}},
  \bibinfo{journal}{Phys. Rev. Lett.} \textbf{\bibinfo{volume}{97}},
  \bibinfo{pages}{021601} (\bibinfo{year}{2006}{\natexlab{a}}),
  \eprint{hep-ph/0601236}.

\bibitem[{\citenamefont{Cohen and Glashow}(2006{\natexlab{b}})}]{Cohen:2006ir}
\bibinfo{author}{\bibfnamefont{A.~G.} \bibnamefont{Cohen}} \bibnamefont{and}
  \bibinfo{author}{\bibfnamefont{S.~L.} \bibnamefont{Glashow}}
  (\bibinfo{year}{2006}{\natexlab{b}}), \eprint{hep-ph/0605036}.

\bibitem[{\citenamefont{Dunn and Mehen}(2006)}]{Dunn:2006xk}
\bibinfo{author}{\bibfnamefont{A.}~\bibnamefont{Dunn}} \bibnamefont{and}
  \bibinfo{author}{\bibfnamefont{T.}~\bibnamefont{Mehen}}
  (\bibinfo{year}{2006}), \eprint{hep-ph/0610202}.

\bibitem[{\citenamefont{Nayak and Jain}(2017)}]{Nayak:2016zed}
\bibinfo{author}{\bibfnamefont{A.~C.} \bibnamefont{Nayak}} \bibnamefont{and}
  \bibinfo{author}{\bibfnamefont{P.}~\bibnamefont{Jain}},
  \bibinfo{journal}{Phys. Rev.} \textbf{\bibinfo{volume}{D96}},
  \bibinfo{pages}{075020} (\bibinfo{year}{2017}), \eprint{1610.01826}.

\bibitem[{\citenamefont{Ahluwalia and Lee}(2017)}]{Ahluwalia:2017rsu}
\bibinfo{author}{\bibfnamefont{D.~V.} \bibnamefont{Ahluwalia}}
  \bibnamefont{and} \bibinfo{author}{\bibfnamefont{C.-Y.} \bibnamefont{Lee}},
  \bibinfo{journal}{EPL} \textbf{\bibinfo{volume}{119}}, \bibinfo{pages}{61001}
  (\bibinfo{year}{2017}), \eprint{1705.09066}.

\bibitem[{\citenamefont{Fan et~al.}(2007)\citenamefont{Fan, Goldberger, and
  Skiba}}]{Fan:2006nd}
\bibinfo{author}{\bibfnamefont{J.}~\bibnamefont{Fan}},
  \bibinfo{author}{\bibfnamefont{W.~D.} \bibnamefont{Goldberger}},
  \bibnamefont{and} \bibinfo{author}{\bibfnamefont{W.}~\bibnamefont{Skiba}},
  \bibinfo{journal}{Phys. Lett.} \textbf{\bibinfo{volume}{B649}},
  \bibinfo{pages}{186} (\bibinfo{year}{2007}), \eprint{hep-ph/0611049}.

\bibitem[{\citenamefont{Marciano and Sanda}(1977)}]{Marciano:1977wx}
\bibinfo{author}{\bibfnamefont{W.~J.} \bibnamefont{Marciano}} \bibnamefont{and}
  \bibinfo{author}{\bibfnamefont{A.~I.} \bibnamefont{Sanda}},
  \bibinfo{journal}{Phys. Lett.} \textbf{\bibinfo{volume}{67B}},
  \bibinfo{pages}{303} (\bibinfo{year}{1977}).

\bibitem[{\citenamefont{Lee and Shrock}(1977)}]{Lee:1977tib}
\bibinfo{author}{\bibfnamefont{B.~W.} \bibnamefont{Lee}} \bibnamefont{and}
  \bibinfo{author}{\bibfnamefont{R.~E.} \bibnamefont{Shrock}},
  \bibinfo{journal}{Phys. Rev.} \textbf{\bibinfo{volume}{D16}},
  \bibinfo{pages}{1444} (\bibinfo{year}{1977}).

\bibitem[{\citenamefont{Pal and Wolfenstein}(1982)}]{Pal:1981rm}
\bibinfo{author}{\bibfnamefont{P.~B.} \bibnamefont{Pal}} \bibnamefont{and}
  \bibinfo{author}{\bibfnamefont{L.}~\bibnamefont{Wolfenstein}},
  \bibinfo{journal}{Phys. Rev.} \textbf{\bibinfo{volume}{D25}},
  \bibinfo{pages}{766} (\bibinfo{year}{1982}).

\bibitem[{\citenamefont{Schechter and Valle}(1981)}]{Schechter:1981hw}
\bibinfo{author}{\bibfnamefont{J.}~\bibnamefont{Schechter}} \bibnamefont{and}
  \bibinfo{author}{\bibfnamefont{J.~W.~F.} \bibnamefont{Valle}},
  \bibinfo{journal}{Phys. Rev.} \textbf{\bibinfo{volume}{D24}},
  \bibinfo{pages}{1883} (\bibinfo{year}{1981}), \bibinfo{note}{[Erratum: Phys.
  Rev.D25,283(1982)]}.

\bibitem[{\citenamefont{Shrock}(1982)}]{Shrock:1982sc}
\bibinfo{author}{\bibfnamefont{R.~E.} \bibnamefont{Shrock}},
  \bibinfo{journal}{Nucl. Phys.} \textbf{\bibinfo{volume}{B206}},
  \bibinfo{pages}{359} (\bibinfo{year}{1982}).

\bibitem[{\citenamefont{Masood}(1993)}]{Masood:1992dg}
\bibinfo{author}{\bibfnamefont{S.~S.} \bibnamefont{Masood}},
  \bibinfo{journal}{Phys. Rev.} \textbf{\bibinfo{volume}{D48}},
  \bibinfo{pages}{3250} (\bibinfo{year}{1993}).

\bibitem[{\citenamefont{Masood}(1995)}]{Masood:1995aq}
\bibinfo{author}{\bibfnamefont{S.~S.} \bibnamefont{Masood}},
  \bibinfo{journal}{Astropart. Phys.} \textbf{\bibinfo{volume}{4}},
  \bibinfo{pages}{189} (\bibinfo{year}{1995}).

\bibitem[{\citenamefont{Dvornikov and Studenikin}(2004)}]{Dvornikov:2004sj}
\bibinfo{author}{\bibfnamefont{M.~S.} \bibnamefont{Dvornikov}}
  \bibnamefont{and} \bibinfo{author}{\bibfnamefont{A.~I.}
  \bibnamefont{Studenikin}}, \bibinfo{journal}{J. Exp. Theor. Phys.}
  \textbf{\bibinfo{volume}{99}}, \bibinfo{pages}{254} (\bibinfo{year}{2004}),
  \eprint{hep-ph/0411085}.

\bibitem[{\citenamefont{Bell et~al.}(2005)\citenamefont{Bell, Cirigliano,
  Ramsey-Musolf, Vogel, and Wise}}]{Bell:2005kz}
\bibinfo{author}{\bibfnamefont{N.~F.} \bibnamefont{Bell}},
  \bibinfo{author}{\bibfnamefont{V.}~\bibnamefont{Cirigliano}},
  \bibinfo{author}{\bibfnamefont{M.~J.} \bibnamefont{Ramsey-Musolf}},
  \bibinfo{author}{\bibfnamefont{P.}~\bibnamefont{Vogel}}, \bibnamefont{and}
  \bibinfo{author}{\bibfnamefont{M.~B.} \bibnamefont{Wise}},
  \bibinfo{journal}{Phys. Rev. Lett.} \textbf{\bibinfo{volume}{95}},
  \bibinfo{pages}{151802} (\bibinfo{year}{2005}), \eprint{hep-ph/0504134}.

\bibitem[{\citenamefont{Hu et~al.}(2018)}]{Hu:2018lwv}
\bibinfo{author}{\bibfnamefont{J.}~\bibnamefont{Hu}} \bibnamefont{et~al.}
  (\bibinfo{year}{2018}), \eprint{1812.11480}.

\bibitem[{\citenamefont{Tiengo et~al.}(2013)}]{Tiengo:2013gsa}
\bibinfo{author}{\bibfnamefont{A.}~\bibnamefont{Tiengo}} \bibnamefont{et~al.},
  \bibinfo{journal}{Nature} \textbf{\bibinfo{volume}{500}},
  \bibinfo{pages}{312} (\bibinfo{year}{2013}), \eprint{1308.4987}.

\bibitem[{\citenamefont{Barr et~al.}(1990)\citenamefont{Barr, Freire, and
  Zee}}]{PhysRevLett.65.2626}
\bibinfo{author}{\bibfnamefont{S.~M.} \bibnamefont{Barr}},
  \bibinfo{author}{\bibfnamefont{E.~M.} \bibnamefont{Freire}},
  \bibnamefont{and} \bibinfo{author}{\bibfnamefont{A.}~\bibnamefont{Zee}},
  \bibinfo{journal}{Phys. Rev. Lett.} \textbf{\bibinfo{volume}{65}},
  \bibinfo{pages}{2626} (\bibinfo{year}{1990}).

\bibitem[{\citenamefont{Babu et~al.}(1992)\citenamefont{Babu, Chang, Keung, and
  Phillips}}]{PhysRevD.46.2268}
\bibinfo{author}{\bibfnamefont{K.~S.} \bibnamefont{Babu}},
  \bibinfo{author}{\bibfnamefont{D.}~\bibnamefont{Chang}},
  \bibinfo{author}{\bibfnamefont{W.~Y.} \bibnamefont{Keung}}, \bibnamefont{and}
  \bibinfo{author}{\bibfnamefont{I.}~\bibnamefont{Phillips}},
  \bibinfo{journal}{Phys. Rev. D} \textbf{\bibinfo{volume}{46}},
  \bibinfo{pages}{2268} (\bibinfo{year}{1992}).

\bibitem[{\citenamefont{Lindner et~al.}(2017)\citenamefont{Lindner, Radovčić,
  and Welter}}]{Lindner:2017uvt}
\bibinfo{author}{\bibfnamefont{M.}~\bibnamefont{Lindner}},
  \bibinfo{author}{\bibfnamefont{B.}~\bibnamefont{Radovčić}},
  \bibnamefont{and} \bibinfo{author}{\bibfnamefont{J.}~\bibnamefont{Welter}},
  \bibinfo{journal}{JHEP} \textbf{\bibinfo{volume}{07}}, \bibinfo{pages}{139}
  (\bibinfo{year}{2017}), \eprint{1706.02555}.

\end{thebibliography}
\end{document}